\documentclass[fleqn,usenatbib]{rasti}

\usepackage{newtxtext,newtxmath}

\usepackage[T1]{fontenc}

\DeclareRobustCommand{\VAN}[3]{#2}
\let\VANthebibliography\thebibliography
\def\thebibliography{\DeclareRobustCommand{\VAN}[3]{##3}\VANthebibliography}

\usepackage{graphicx}	
\usepackage{amsmath}	
\usepackage{lineno}
\usepackage{booktabs} 
\usepackage{multirow} 

\title[MauveSim: the instrument simulator software for the Blue Skies Space Mauve satellite]{MauveSim: the instrument simulator software for the Blue Skies Space Mauve satellite}

\author[A. Saba et al.]{
Arianna Saba,$^{1}$\thanks{E-mail: arianna@bssl.space (AS)}
Fabio Favata,$^{1,2,3}$
Giorgio Savini,$^{1,4}$
Giovanna Tinetti,$^{1,5}$
Lawrence Bradley,$^{1}$
\newauthor
Ian Stotesbury,$^{1}$ 
Marcell Tessenyi$^{1}$
\\
$^{1}$Blue Skies Space Ltd., 69 Wilson Street, London, EC2A 2BB, UK\\
$^{2}$INAF - Osservatorio Astronomico di Palermo, Piazza del Parlamento, 1, 90134 Palermo, Italy\\
$^{3}$Department of Physics, Imperial College London, Exhibition Road, London SW7 2AZ, United Kingdom\\
$^{4}$Department of Physics and Astronomy, University College London, Gower Street, WC1E 6BT London, United Kingdom\\
$^{5}$Department of Physics, King's College London, Strand, WC2R 2LS London, United Kingdom\\
}

\date{Accepted XXX. Received YYY; in original form ZZZ}

\pubyear{\the\year{}}

\begin{document}
\label{firstpage}
\pagerange{\pageref{firstpage}--\pageref{lastpage}}
\maketitle

\begin{abstract}
We present MauveSim, the instrument simulator software for Mauve, the latest mission from Blue Skies Space dedicated to time-domain stellar astronomy. MauveSim functions as an end-to-end simulator, employing the most up-to-date knowledge of the instrument's performance and characteristics that will be reviewed and updated after commissioning. The software accepts a stellar spectrum--either observed or synthetic--as input and produces a simulated observation. The tool thus enables the assessment of various scientific objectives, as well as determining limiting magnitudes and conducting signal-to-noise (S/N) analyses. The results of MauveSim have been validated against instrument performance data from extensive ground testing campaigns, ensuring that the software reflects the most up-to-date understanding of the payload performance. Accessible to all scientists involved in the mission, MauveSim serves as a crucial tool for target selection and observation planning.
\end{abstract}




\section{Introduction}
Born in the 1960s, mid and far ultraviolet (UV) astronomy coincided with the advent of space astronomy, as Earth's atmosphere absorbs light at these wavelengths. Despite the great advancement of space astronomy in recent decades, UV observatories remain scarce and are often oversubscribed (e.g. the Hubble Space Telescope). In contrast, our understanding of astronomical objects and transient phenomena in the visible and near-infrared has expanded at an unprecedented pace driven by the large number of facilities deployed on the ground and in space. Time-domain astronomy, which focuses on the variable emission of sources, has been identified as a key priority for the next decade, as emphasised by the “Decadal Survey” outlining priorities for the USA research community \citep{2021pdaa.book.....N}. A small space telescope capable of tracking variability across a broad wavelength range, including portions of the UV spectrum, will fill a crucial niche and enable unique scientific discoveries in this field \citep{egan2022colorado, 2022apra.prop..121I}. 

In this landscape, Blue Skies Space Ltd. (BSSL) has conceived Mauve, a 16U smallsat designed to address the current gap in UV data availability. The mission is dedicated to time-domain spectrophotometry in the UV and Visible (UV-Vis) spectral range. Throughout its lifetime, Mauve will obtain time series of low-resolution spectra at $\mathscr{R}$ = 20-65 over a broad spectral range (200-700 nm), enabling the investigation of astrophysical phenomena for which so far only fragmentary data have been collected. Even a small satellite like Mauve will be able to provide much needed coverage in the UV and optical bands, which can complement existing spectroscopic stellar data at longer wavelengths. Mauve can effectively survey stellar targets in the UV and visible, enabling detailed characterisation of stellar activity and variability, key for interpreting exoplanet spectroscopic observations. The capabilities of Mauve in monitoring stellar targets are especially valuable in the current space observatories landscape, given the James Webb Space Telescope's lack of sensitivity below 600 nm \citep{2016ApJ...817...17G}. Additionally, Mauve will be capable of obtaining near-ultraviolet spectra and constrain the variability of stars identified as high-priority targets for upcoming exoplanet-dedicated missions such as Ariel \citep{2018ExA....46..135T, 2021arXiv210404824T} and Twinkle \citep{2019ExA....47...29E, 2022SPIE12180E..33S} and the future Habitable Worlds Observatory \citep{2024arXiv240212414M}.

Mauve’s design philosophy differs from traditional space science telescopes, which start from detailed science specifications around which a dedicated instrument is designed and built. While this approach results in high quality, high-performance instruments, it has traditionally come at significant financial cost. In keeping with BSSL’s philosophy \citep{2020NatAs...4.1017A}, Mauve aims to fly commercial off-the-shelf (COTS) systems, resulting in a cost orders of magnitude lower than custom-designed instruments, albeit, unavoidably, with some performance compromises. In the case of Mauve, the key compromise is the high dark current from the detector due to the lack of a dedicated cryogenic system. Mauve is therefore optimised for long-term monitoring of the broad-band spectrophotometric variability of relatively bright sources, for which the high dark current is not a significant limiting factor.

The spacecraft will be launched in a low Earth Sun-synchronous orbit at 500 km of altitude, with an orbital period of approximately 95 minutes. As a result, Mauve will intersect regularly with the South Atlantic Anomaly (SAA). BSSL's intention is to use the commissioning phase to qualify the performance during SAA crossing and determine if there is an impact on the science measurements. If needed, BSSL will determine a flux density cut-off above which to cease payload operations and pause the scheduled observations.

In this paper, we present MauveSim, the instrument simulator developed by BSSL for the Mauve mission. Instrument simulators are essential tools for mission preparation. They are designed to replicate the performance of astronomical instruments, incorporating information from design choices and the best understanding of the instrument's characteristics and performance evaluated during ground testing campaigns \citep{simspecdocs, 2010PASP..122..947D, moonsim}. As such, an instrument simulator naturally develops over time, as the understanding of the instrument evolves, and it is updated in flight when the final calibration is achieved. Simulators also enable scientists to test, propose in an informed way, and analyse observations at all stages--before, during, and after data acquisition--providing insights into the influence of instrumental effects on the data.
Moreover, simulators are vital for optimising observation strategies and validating data reduction pipelines.

MauveSim is written entirely in Python and accessible on Stardrive, a web-based platform developed by BSSL that integrates simulation results, data management, and collaboration tools in a unified environment. Stardrive is available for all scientists participating in the Mauve mission. At BSSL, we are supporting MauveSim as an actively maintained tool within the Stardrive platform, with regular updates driven by mission requirements and feedback from the Mauve science consortium. The instrument simulator is available to Mauve users through a secure, web-based interface as part of our mission planning and data management suite. We also offer ad-hoc access upon request. The software is maintained under version control and the software version is indicated in the web-interface. Our long-term strategy additionally includes ongoing support and continuous improvements to ensure MauveSim remains reliable, scientifically accurate, and aligned with the evolving needs of the mission.

This paper is organised as follows. Section~\ref{sec:payload} presents the Mauve's payload design and the characteristics of its components. More details on the simulator validation activities are available in Section~\ref{sec:performance},  where we also describe the limitations we encountered during the testing campaign when trying to characterise the payload as accurately as possible. The description of the algorithm is presented in Section~\ref{sec:workflow}, which explains in detail the assumptions, input data, implementation and output of each step in the workflow. Finally, Section~\ref{sec:science_cases} presents the application of the instrument simulator to one of the Mauve "flagship" science cases, showing how the software can be used to understand limiting magnitudes, and thus perform science definition.

\section{Payload components}
\label{sec:payload}
\begin{figure}
    \centering
    \includegraphics[width=1\linewidth]{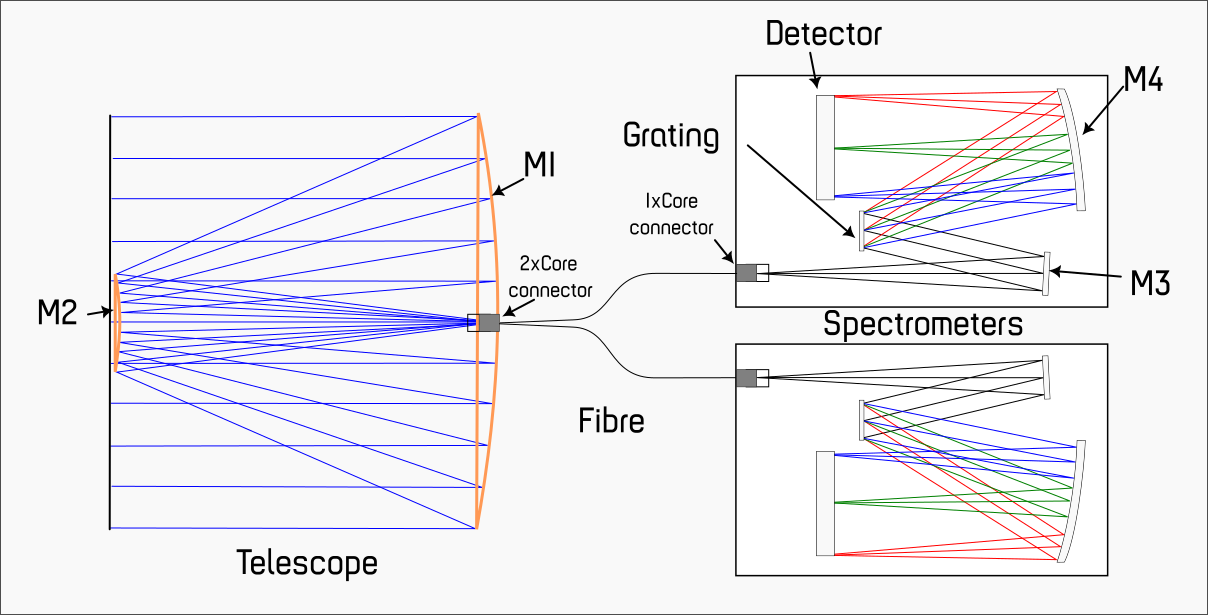}
    \caption{Illustration of Mauve optical diagram (not to scale).}
    \label{fig:optical_design}
\end{figure}
\begin{table}
\centering
\caption{Mauve's spectrometer, detector and telescope specifications.}
\label{tab:MAUVE_spectrometer_detector}
\begin{tabular}{|c|c|}
  \hline
  \multicolumn{2}{|c|}{\textbf{Spectrometer Key Properties}} \\
  \hline
  Grating & 600 lines/mm \\
  AD converter & 16-bit, 6 MHz \\
  Dimensions & 105$\times$80$\times$20 mm \\
  Weight & 277.5 g \\
  \hline
  \multicolumn{2}{|c|}{\textbf{Detector Specifications}} \\
  \hline
  Detector type & HAM S11639, CMOS linear array \\
  Wavelength coverage & 200-700 nm \\
  Wavelength range per pixel ($\Delta x_{\lambda}$) &  0.25-0.30 nm \\
  Array Size & 2048$\times$1 pix \\
  Pixel Size & 14$\times$200 $\mu$m \\
  Read Noise & 12 $\text{e}^{-}$ rms 
  \\
  Dark current & 445 counts/pix $\cdot$ s (at 21$^{\circ}$C) \\
  \hline
  \multicolumn{2}{|c|}{\textbf{Telescope Parameters}} \\
  \hline
  Field of View & 47.46'' radius \\
  Collecting area ($A$) & 122 cm$^2$ \\
  \hline
\end{tabular}
\end{table}
The Mauve payload consists of three main components: a telescope, an optical fibre and a spectrometer. The payload has been designed with redundancy in mind, so the current baseline includes the provision of two fibre+spectrometer units to cope with a potential failure of the primary unit. 
The telescope and optical components are housed separately from the spacecraft platform equipment to minimise thermal variations throughout the orbit and reduce thermo-elastic distortions. The design, developed by C3S and ISISPACE, leverages heritage components where feasible and utilises the product lines and supply chains of both companies \citep{iantechnicalsheet}.

The selected telescope is a Cassegrain telescope of 13 cm aperture originally designed for space-based optical communications by MediaLario, Italy. The telescope has already undergone a wide range of optical, thermal and vibration tests as part of qualification campaigns on other satellites and required only two minor modifications to be compatible with the Mauve satellite: 
\begin{itemize}
    \item As the telescope was originally designed for laser optical communication, the standard coating is optimised for 1550 nm. The current anti-reflection (AR) coating has been replaced with a coating designed to improve the signal in the UV range. 
    \item To mount the fibre between the telescope and the spectrometer, MediaLario modified the mount of the telescope to allow the fibre to be placed at the telescope focus. The fibre will exit from the back of the primary mirror and connect to the spectrometer, aligning its output with the spectrometer aperture. The payload optical design is provided in Fig.~\ref{fig:optical_design}.
\end{itemize}

The spectrometer is produced by Avantes, Netherlands. The unit is a high-quality, COTS spectrometer used across biomedicine, semiconductor coating detection and other terrestrial applications. The spectrometer houses a diffraction grating, a detector collection lens (DCL), and a window to limit contamination into the spectrometer through the connector port. The spectrometer operates from 200 to 700 nm across a single channel, illuminating a CMOS line array detector from Hamamatsu \citep{hamadetector}. Spectrometer and detector characteristics are given in Table~\ref{tab:MAUVE_spectrometer_detector}.

\section{Instrument performance}
\label{sec:performance}
\begin{table}
    \centering
    \caption{Payload performance specifications.}
    \begin{tabular}{c|c}
    \hline
    Parameter & Value \\
    \hline
    Detector temperature & $21^{\circ}\mathrm{C}$ \\
    Dark current & 445 counts/s \\
    Light loss across the system & 34\% \\
    \hline
    \end{tabular}
    \label{tab:performance}
\end{table}
\begin{figure}
    \centering
    \includegraphics[width=1\linewidth]{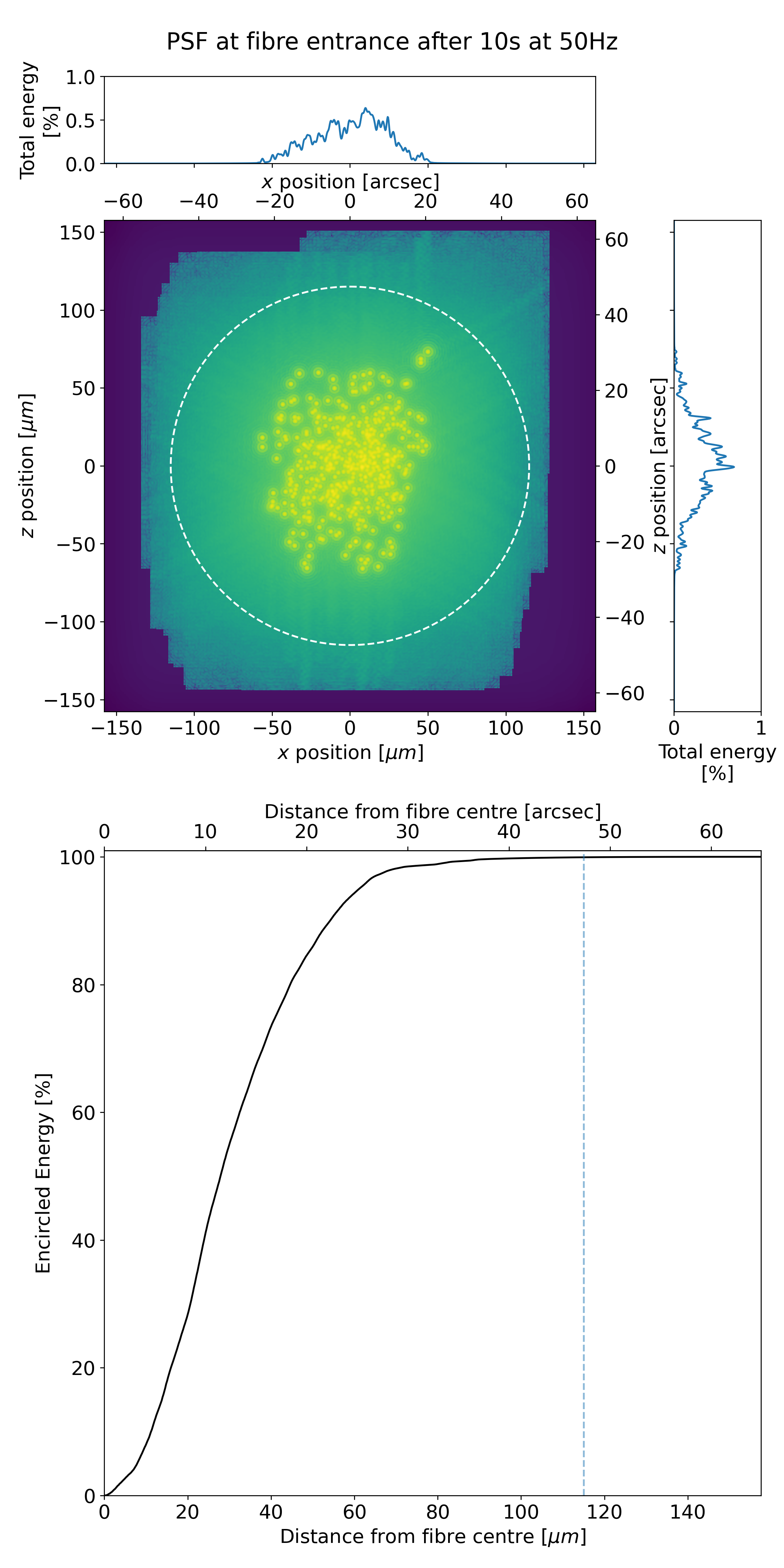}
    \caption{Mauve PSF at the fibre entrance during a 10 s integration. The PSF is plotted on a logarithmic scale and has been normalised such that its total sum is 1. The fibre radius is 115 $\mu$m, with a plate scale of 0.412"/$\mu$m.}
    \label{fig:jitter}
\end{figure}
We evaluated Mauve's performance through a series of laboratory measurements of the telescope, fibre, and spectrometer, obtained during several testing campaigns at MediaLario. During the functional tests, the payload's performance was assessed at multiple stages along the optical chain. Measurements were performed using two narrow-band filters: one in the ultraviolet region at 248 nm and another in the visible wavelengths at 636 nm. To validate the results, measured signal outputs were compared against expected values based on the efficiency curves of each payload component, as provided by the manufacturers. A consistent signal level of approximately 34\% below the expected signal value was observed across all tests, indicating a lower efficiency of the payload components. While we don't expect this to impact results significantly, we note that the laboratory tests we performed had limitations in producing a perfectly collimated beam, potentially impacting the coupling to the fibre. Furthermore, the estimation of UV performance is less accurate than that of visible wavelength performance, due to wavelength dependence in scattering properties of the telescope mirrors compared to band integrated measurements of efficiency.

The conservative thermal model supplied by the manufacturers assumes a detector operating temperature of $21^{\circ}\mathrm{C}$. While dedicated spacecraft instruments typically operate at cooler temperatures, Mauve’s COTS spectrograph is subject to higher temperatures due to design constraints. Ongoing work with the manufacturing partners aims to lower the detector temperature to a target of \(13^{\circ}\mathrm{C}\). A refined thermal analysis, including a margin approach, suggests that this could halve the dark current and significantly improve instrument performance.

The example science simulations in Sec.~\ref{sec:science_cases} are based on the conservative thermal model while the system's light loss is the one measured in the lab. These parameters are presented in Table~\ref{tab:performance}.


A third factor influencing and potentially limiting scientific performance is relying on payload data for pointing, rather than solely using star tracker information. This approach will be essential to compensate for the expected thermal distortion caused by the spacecraft crossing the Earth’s terminator during each orbit. 
Our concept consists in determining the fraction of time that the source’s light is actually collected by the fibre, as opposed to being lost when the source falls outside the fibre. The approach taken in defining the diameter of the fibre and thus the Field of View (FoV) has been to use the high-frequency pointing performance of the spacecraft. This has been modelled by ISISPACE using the Moments of Inertia (MoI) calculations of the mature mechanical model and the Attitude Determination and Control System (ADCS) controller and actuators in the system. The ADCS implements an attitude estimation and control loop, making use of high-performance gyros and star trackers to provide the measurements, and reaction wheels and magnetorquers for actuation. The relative pointing error (RPE) of the system is estimated to be approximately 12'' across a 10 s window. This information has been incorporated into a pointing model for the system, enabling time-domain jitter analysis. Our estimates indicate that jitter-induced noise is negligible. The fibre diameter is 230 $\mu$m, corresponding to a half-cone FoV of approximately 47''. As our detector is a linear array, pointing jitter will affect the resolving power but not the signal to noise ratio as long as all the light from the target star enters the fibre optic at the focal point of the telescope. The diameter of the fibre optic is much larger than the dispersion caused by pointing jitter. A "Payload in the Loop" (PITL) control algorithm was developed by BSSL to address low-frequency pointing distortions. These distortions occur when the satellite crosses the Earth's terminator zone \citep{lawrence2025}. This algorithm is designed to provide corrections to the ADCS control loop so as to minimise light loss and keep all the light from the target star centred in the fibre optic at the telescope focal plane. Simulations with the PITL indicate that light loss due to pointing errors will be less than 5\%. In Fig.~\ref{fig:jitter} we present the evolution of the pointing jitter across a 10 s window and how this affects the point spread function (PSF) shape at fibre entrance. However, the instrument simulator operates independently of the PITL’s performance. The simulations presented here are representative of periods when the satellite is thermally stable, maintains steady pointing, and can successfully lock onto and track the target source selected. How often and how much the jitter distortions will affect the science data can only be accurately estimated in orbit.

\section{Workflow structure}
\label{sec:workflow}
MauveSim produces simulated Mauve observations by applying a series of steps. These can be grouped in 4 broad categories.
\begin{enumerate}
    \item \textit{Preparing the input spectrum}: Downsampling the initial number of data points, scaling the visible magnitude of the target source, including zodiacal background contamination.
    \item \textit{Convolving the spectrum to the resolution of the instrument}: Applying the instrument line function (ILF) to match the input data to the spectral resolution of the spectrometer.
    \item \textit{Converting flux into counts}: Considering the efficiencies of the payload components, the telescope aperture, detector quantum efficiency, wavelength coverage per pixel, exposure time and conversion factors.
    \item \textit{Adding astrophysical and instrumental noise sources}: Including photon noise, dark current and bias.
\end{enumerate}

\subsection{Preparing the input spectrum} 
In the MauveSim interface we provide an integrated library of stellar spectra that enables the users to select appropriate inputs for their simulations. These stellar spectra are sourced from the HST/STIS Stellar Spectral Library\footnote{\url{http://astro.wsu.edu/hststarlib/}} \citep{2023ApJS..266...41P} which provides more than 500 calibrated stellar spectra across the 170-1000 nm wavelength region at $\mathscr{R}$ $\sim$ 1000. Spectra from the STIS library are given in units of \AA \ for the wavelength and $\text{erg} \, \text{s}^{-1} \, \text{cm}^{-2} \, \text{\AA}^{-1}$ \ for the spectral flux density.

Alternatively, users may upload their own spectrum via the simulator interface, provided the input data adheres to specified guidelines. The required input file must be in ASCII format, with a .txt or .dat extension. The file must contain two numerical columns separated by spaces, tabs or commas. Column 1 must contain the wavelength array in units of nm, while the second column must contain the spectral flux density array in units of $\text{erg} \, \text{s}^{-1} \, \text{cm}^{-2} \, \text{\AA}^{-1}$. The spectrum must span at least 180–750 nm, which is the detector's operational spectral band. While the detector can capture light beyond the spectrometer’s wavelength range of 200-700, the payload’s efficiency outside of this range is generally too poor to produce scientifically useful data. To account for this, the software restricts the output to the 200–700 nm range.

\label{sec:preparing_input_spectrum}
\begin{figure}
    \centering
    \includegraphics[width=\columnwidth]{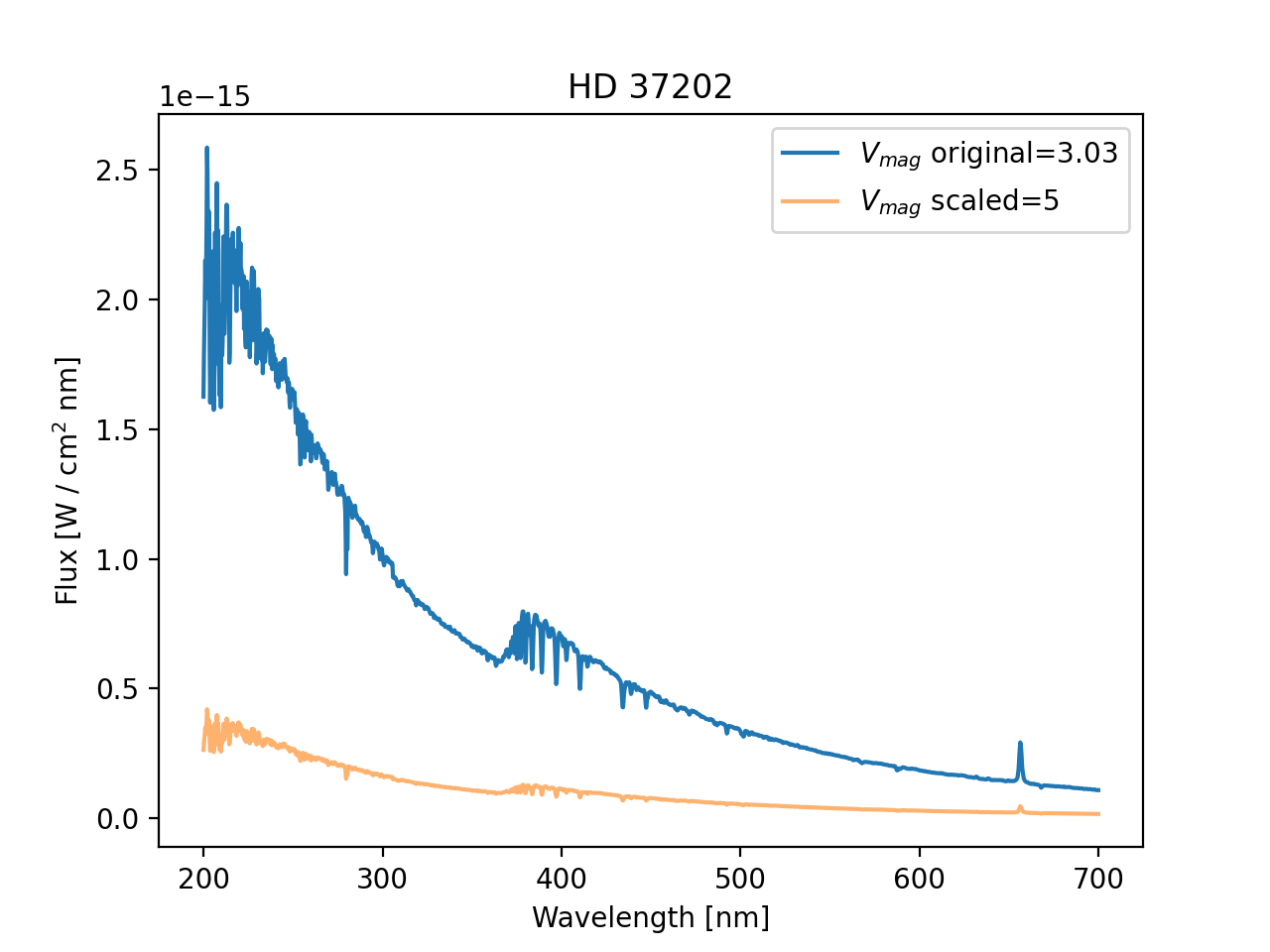}
    \caption{Original spectrum of HD 37202 (V$_{\rm mag}$=3.03) compared to its rescaled version at V$_{\rm mag}$=5. This example demonstrates the instrument simulator's capability to modify the input flux to a user's specified magnitude.}
    \label{fig:MAUVE_example}
\end{figure}
Once the user selects the input spectrum, this is downsampled if it contains more than 2048 data points, the number of detector pixels in the spectral direction. This interpolation is necessary to ensure that each wavelength data point is consistent with the association of the detector array with the range created by the combination of the grating plus order sorting filter.

To illustrate the simulation process, we use the example of $\zeta$ Tauri (HD 37202), a Be star with a visual magnitude of 3.03. Classical Be (CBe) stars are B-type stars located on the upper main sequence that are characterized by prominent emission features—particularly in the H$\alpha$ line—alongside near- and mid-infrared excess emission beyond 1$\mu$m and a continuum polarization of roughly one percent. CBe stars are currently understood to host a thin, gaseous, equatorial disk, formed from material ejected by the star, where both the emission lines and infrared excess originate through recombination processes \citep{2013A&ARv..21...69R}. Despite extensive study, the exact mechanism responsible for disk formation remains unsolved. Mauve offers an opportunity to conduct an optical and ultraviolet spectral energy distribution (SED) survey of hundreds of later-type Be stars, potentially offering new insights into the physical processes governing these enigmatic systems.

The software includes an option to simulate the input spectrum at a different magnitude. This is achieved by scaling the flux according to
\begin{equation}
    F = F_0 \cdot 10^{-\Delta m / 2.5} \ ,
    \label{eq:magnitude}
\end{equation}
where $F_0$ is the original flux and $\Delta m$ = $m_{\text{final}}$ - $m_{\text{initial}}$, i.e. the difference between the final wanted magnitude (e.g. $V_\text{mag}$ = 5) and the initial magnitude (in this case $V_\text{mag}$ = 3.03). To illustrate the magnitude rescaling option, Fig.~\ref{fig:MAUVE_example} displays the original spectrum of HD 37202 retrieved from the STIS stellar database, alongside its version scaled to a visual magnitude of 5.

\begin{table*}
\centering
\caption{Zodiacal sky background ($V_{\text{mag}}$ $\text{arcsec}^{-2}$) as a function of helio-ecliptic coordinates. Table from the \href{https://hst-docs.stsci.edu/cosihb/chapter-7-exposure-time-calculator-etc/7-4-detector-and-sky-backgrounds\#id-7.4DetectorandSkyBackgrounds-Figure7.2}{COS Instrument Handbook.}}
\label{tab:zodi}
\begin{tabular}{|c|c|c|c|c|c|c|c|c|}
\hline
\multirow{2}{*}{\textbf{Helio-ecliptic Longitude}} & \multicolumn{8}{|c|}{\textbf{Helio-ecliptic Latitude}} \\
\cline{2-9}
& 0$^\circ$ & 15$^\circ$ & 30$^\circ$ & 45$^\circ$ & 60$^\circ$ & 75$^\circ$ & 90$^\circ$ & \\
\hline
180$^\circ$ & 22.1 & 22.4 & 22.7 & 23.0 & 23.2 & 23.4 & 23.3 & \\
\hline
165$^\circ$ & 22.3 & 22.5 & 22.8 & 23.0 & 23.2 & 23.4 & 23.3 & \\
\hline
150$^\circ$ & 22.4 & 22.6 & 22.9 & 23.1 & 23.3 & 23.4 & 23.3 & \\
\hline
135$^\circ$ & 22.4 & 22.6 & 22.9 & 23.2 & 23.3 & 23.4 & 23.3 & \\
\hline
120$^\circ$ & 22.4 & 22.6 & 22.9 & 23.2 & 23.3 & 23.3 & 23.3 & \\
\hline
105$^\circ$ & 22.2 & 22.5 & 22.9 & 23.1 & 23.3 & 23.3 & 23.3 & \\
\hline
90$^\circ$  & 22.0 & 22.3 & 22.7 & 23.0 & 23.2 & 23.3 & 23.3 & \\
\hline
75$^\circ$  & 21.7 & 22.2 & 22.6 & 22.9 & 23.1 & 23.2 & 23.3 & \\
\hline
60$^\circ$  & 21.3 & 21.9 & 22.4 & 22.7 & 23.0 & 23.2 & 23.3 & \\
\hline
45$^\circ$  & SA   & SA   & 22.1 & 22.5 & 22.9 & 23.1 & 23.3 & \\
\hline
30$^\circ$  & SA   & SA   & SA   & 22.3 & 22.7 & 23.1 & 23.3 & \\
\hline
15$^\circ$  & SA   & SA   & SA   & SA   & 22.6 & 23.0 & 23.3 & \\
\hline
0$^\circ$   & SA   & SA   & SA   & SA   & 22.6 & 23.0 & 23.3 & \\
\hline
\end{tabular}
\end{table*}

The flux is then converted from units of $\text{erg} \, \text{s}^{-1} \, \text{cm}^{-2} \, \text{\AA}^{-1}$ to $\text{W} \, \text{cm}^{-2} \, \text{nm}^{-1}$ to match the units of the quantities used to convert the flux to counts, e.g. the detector quantum efficiency (QE) which is given in units of A/W.

To calculate the zodiacal noise, we use a table of approximate zodiacal sky background as a function of helio-ecliptic coordinates. The user is prompted to provide the RA, Dec and time of observation (UTC) to check if the target is within the observing zone or in the solar-avoidance zone. The equatorial coordinates RA and Dec are first converted to geocentric ecliptic coordinates. In the geo-ecliptic coordinate system, the zero longitude is defined by the vernal equinox. However, the zodiacal light is a function of the Sun's position. 
By using the input time of observation, the Sun's position from Earth's perspective is calculated at the given time and is transformed to geocentric ecliptic coordinates \citep{1998A&AS..127....1L}. The geocentric ecliptic coordinates are converted to helio-ecliptic coordinates according to:
\begin{equation}
\begin{aligned}
   \lambda_{\text{helio}} &= \lambda_{\text{geo}} - \lambda_{\odot}, \\
   \beta_{\text{helio}} &= \beta_{\text{geo}} \ ,
\end{aligned}
\end{equation}
where $\lambda$ represents longitude, \(\beta\) represents latitude and $\lambda_{\odot}$ represents the Sun geo-ecliptic longitude.
Therefore, the level of zodiacal light is a function of the position of the star in the sky and of the time it is observed (more or less close to the Sun). The helio-ecliptic coordinates thus defined are then assigned the corresponding level of zodiacal background, using the reference Table~\ref{tab:zodi}. 

The zodiacal sky background is given in units of $V_{\text{mag}}$ $\text{arcsec}^{-2}$. To convert it into flux, we assume that the zodiacal background has a spectrum similar to the solar one, which is a reasonable assumption considering the wavelength coverage of Mauve \citep{1998ApJ...508...44K, 2015ExA....40..601P}. On MauveSim we use the spectrum of HD 87646 (a G2V star) and scale its magnitude to
\begin{equation}
    V_{\text{total}} = V_{\text{zodi}} - 2.5 \log_{10}(A_{\text{FoV}}) \ .
    \label{eq:zodi}
\end{equation}
$V_{\text{total}}$ is calculated by taking into account the magnitude of the zodiacal light read off from the table ($V_{\text{zodi}}$) and the field of view of the telescope $(A_{\text{FoV}})$, 47.46 arcsec in radius, corresponding to an area of 7072 arcsec$^2$ $(A_{\text{FoV}})$. Equation~\ref{eq:zodi} comes from the relationship between surface brightness (magnitude per square arcsecond) and the total magnitude over a given area. The surface brightness ($V_{\text{zodi}}$) is the apparent magnitude per square arcsecond. It describes how bright an object appears spread over one square arcsecond of the sky. The total magnitude ($V_{\text{total}}$) is the total apparent magnitude of an object when considering the entire area it covers. When a surface brightness is given in square arcseconds, it is necessary to account for how the flux adds up over that area. The total flux is the surface brightness flux multiplied by the area:
\begin{equation}
    F_{\text{total}} = F_{\text{zodi}} \times A_{\text{FoV}}
\end{equation}

To convert flux to magnitude, we employ the usual relationship:
\begin{equation}
    m_1 - m_2 = -2.5 \log_{10} \left( \frac{F_1}{F_2} \right) \ ,
\end{equation}
where $m_1 = V_{\text{total}}$, $m_2 = V_{\text{zodi}}$, $F_1 = F_{\text{total}}$, and $F_2 = F_{\text{zodi}}$, such that
\begin{equation}
    V_{\text{total}} - V_{\text{zodi}} = -2.5 \log_{10} \left( \frac{F_{\text{total}}}{F_{\text{zodi}}} \right) \ .
\end{equation}
Since $F_{\text{total}}$ = $F_{\text{zodi}} \times A_{\text{FoV}}$, then
\begin{equation}
    V_{\text{total}} - V_{\text{zodi}} = -2.5 \log_{10} (A_{\text{FoV}}) \ .
\end{equation}

\begin{figure}
    \centering
    \includegraphics[width=\columnwidth]{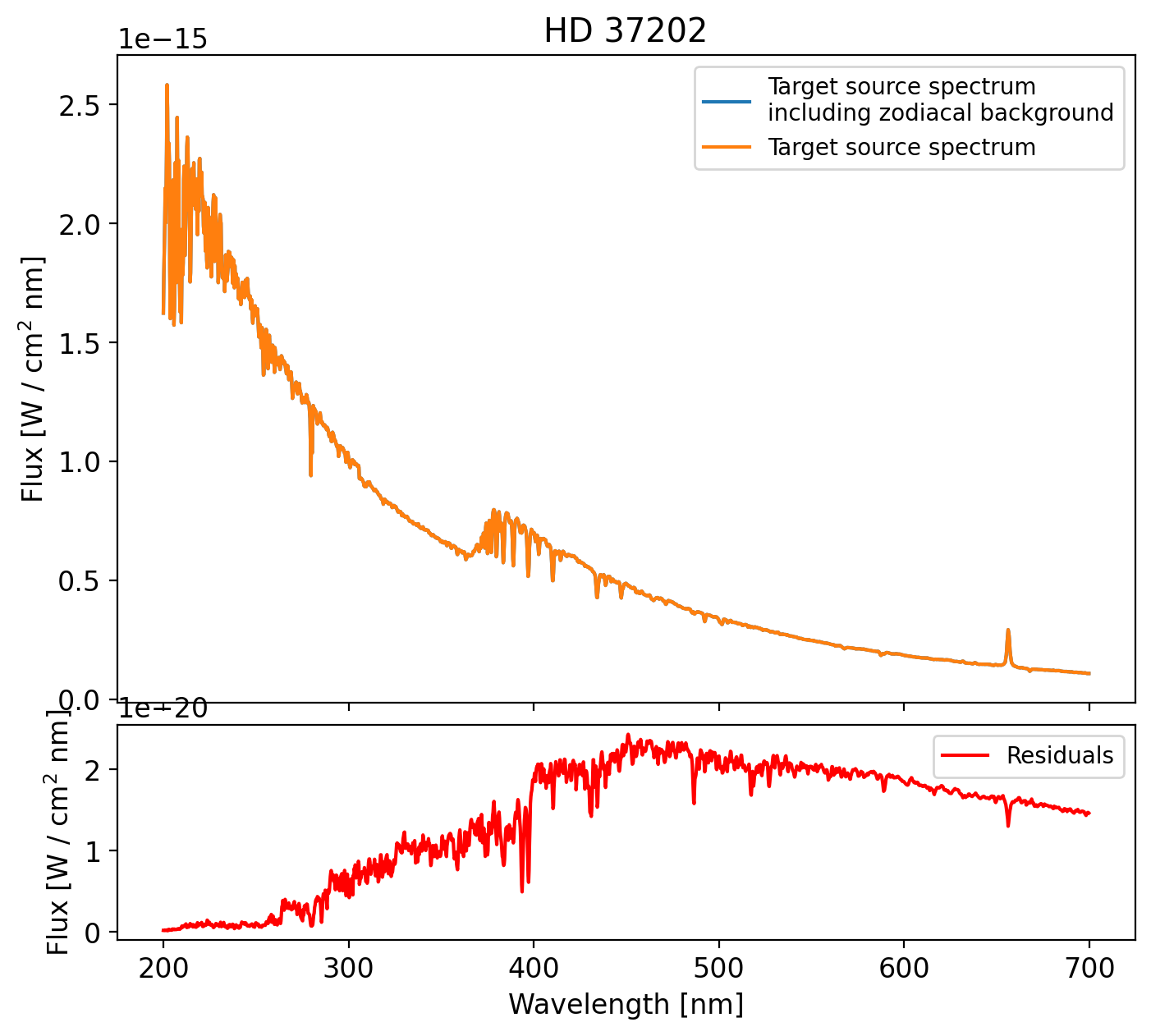}
    \caption{Target stellar spectrum with and without the inclusion of zodiacal background. While the spectra are indistinguishable, the residuals reveal a solar-type spectrum characteristic of zodiacal light.}
    \label{fig:spectrum_with_zodi}
\end{figure}

Fig.~\ref{fig:spectrum_with_zodi} illustrates how the target spectrum is altered by the inclusion of the zodiacal background. While the two spectra appear visually indistinguishable, the residuals reveal that their difference corresponds to a solar-type spectrum with an average $V_{\text{mag}} = 22$.

\subsection{Calculating the source signal}
\begin{figure}
    \centering
    \includegraphics[width=\columnwidth]{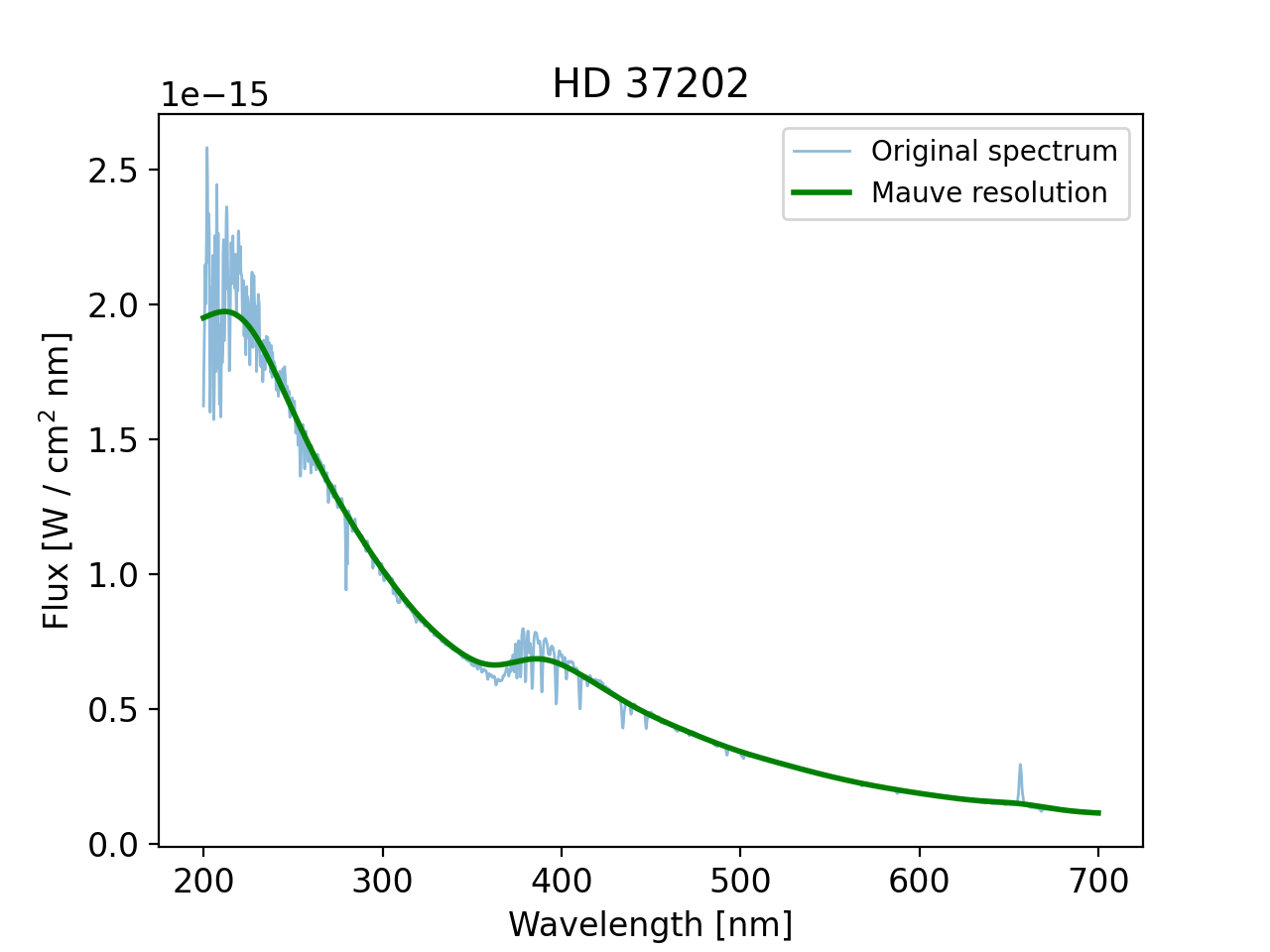}
    \caption{The spectrum of HD 37202 observed with HST/STIS is shown in blue. In green the same spectrum has been convolved to Mauve's instrument line function. The spectrum in green does not represent the flux levels achievable by Mauve.}
    \label{fig:MAUVE_convolved}
\end{figure}
To express the spectrum according to the resolution of the instrument, MauveSim convolves the result of step~\ref{sec:preparing_input_spectrum} with a kernel that downgrades the data to the resolution of Mauve. Here we convolve the stellar spectrum with a kernel representing the instrument line function (ILF).

The convolution \(y(\lambda)\) of a spectrum \(x(\lambda)\) with a kernel \(k(\lambda)\) is given by:
\begin{equation}
    y(\lambda) = (x * k)(\lambda) = \int_{0}^{\infty} x(\lambda') k(\lambda - \lambda') \, d\lambda' \ .
\end{equation}
Here, \(y(\lambda)\) is the resulting convolved spectrum, \(x(\lambda)\) is the stellar spectrum, \(k(\lambda)\) is the kernel function, and \(\lambda\) represents the wavelength. 

The instrument line function, determines how the original spectral data gets broadened. The Mauve ILF features two components: a “top hat” element due to the fibre and a Gaussian component due to the spectrograph. A one-dimensional top hat kernel centred at \( \lambda = 0 \) with width \( 2a \) (i.e., it extends from \( -a \) to \( a \)) is defined as:
\begin{equation}
     k(\lambda) = \begin{cases} 
    \frac{1}{2a} & \text{if } |\lambda| \leq a \\
    0 & \text{if } |\lambda| > a 
    \end{cases}
\end{equation}
Here, \( k(\lambda) \) is the kernel function, and \( a \) is a positive constant representing half the width of the kernel. The kernel is normalised so that the total area under the curve is 1:
\begin{equation}
    \int_{0}^{\infty} k(\lambda) \, d\lambda = 1 \ .
\end{equation}
According to the spectrometer specifications, for a grating of 600 lines/mm and a 500 $\mu$m slit, the resolution is 108 $\text{\AA}$. This is the width of the boxcar kernel.
    
The Gaussian kernel is defined by its characteristic bell-shaped curve
\begin{equation}
k(\lambda) = \frac{1}{\sigma \sqrt{2\pi}} e^{-\frac{\lambda^2}{2\sigma^2}} \ ,
\end{equation}
where the standard deviation $\sigma$ determines the width of the kernel. 

Mauve has constant resolution with wavelength, meaning that the value of $\Delta \lambda$, the smallest resolvable wavelength difference, is the same throughout the entire wavelength range of the instrument. For Mauve, $\Delta \lambda$ $\approx$ 105 $\text{\AA}$, which determines the FWHM of the Gaussian kernel. The top hat kernel and Gaussian kernel are first convolved together, then their convolution is applied to the data. The convolution order does not matter as it is a commutative operation. Fig.~\ref{fig:MAUVE_convolved} displays the spectrum of HD 37202 after convolution with the Mauve instrument line function.

\subsection{Converting flux into counts}
\begin{figure}
\centering
\includegraphics[width=\columnwidth]{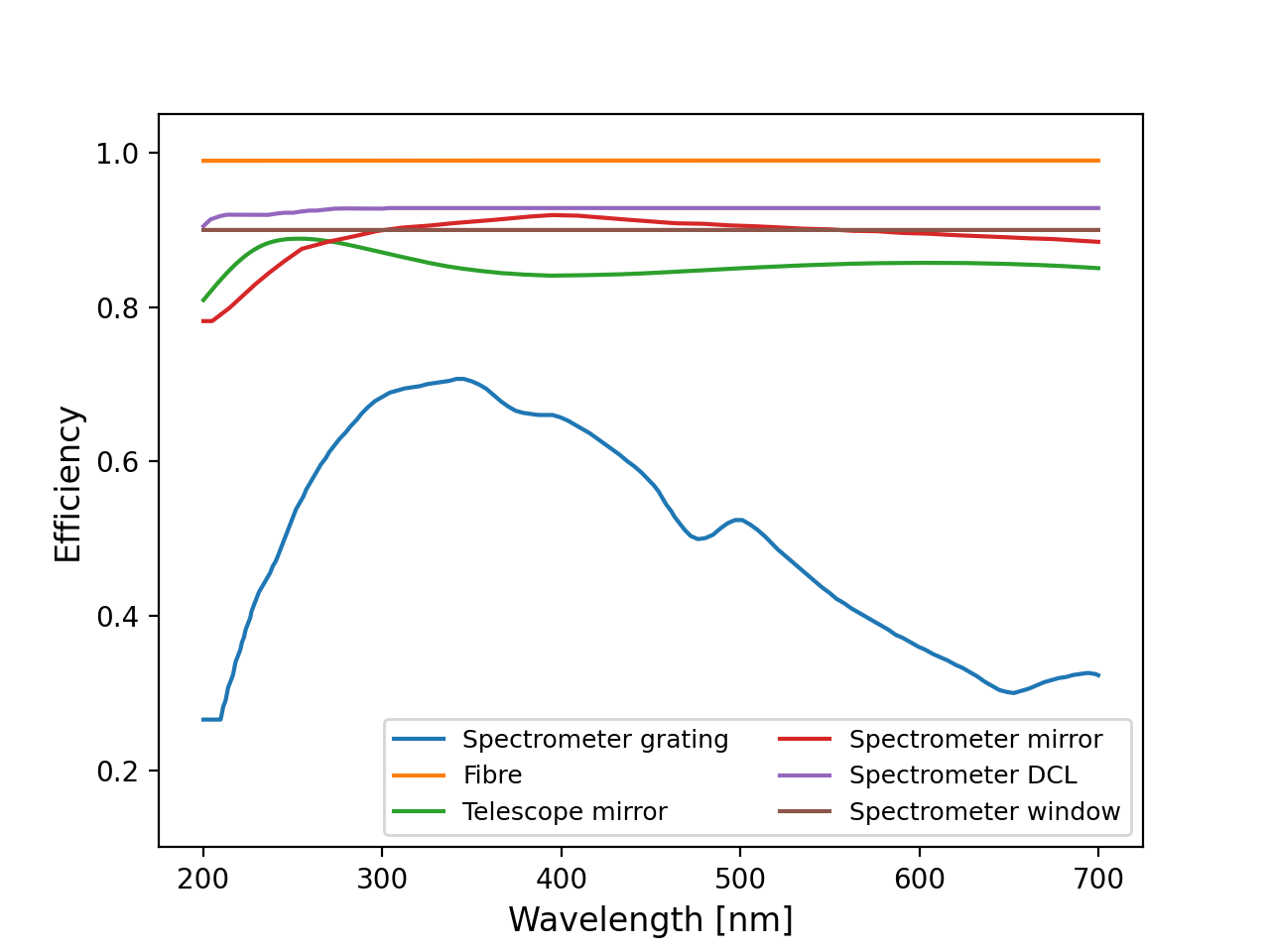}
\caption{Efficiencies of the Mauve payload components.}
\label{fig:efficiencies}
\end{figure}
\begin{figure}
    \centering
    \includegraphics[width=1\linewidth]{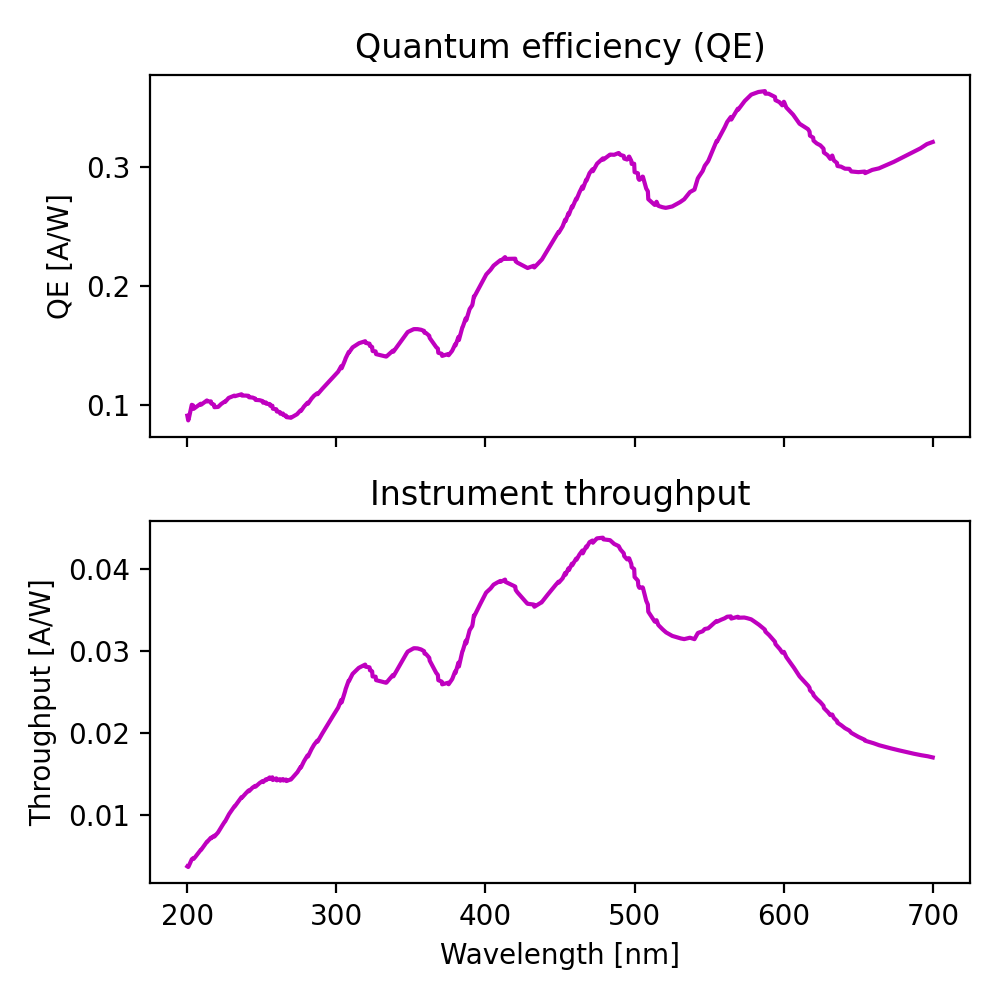}
    \caption{Upper panel: Mauve's detector quantum efficiency. Lower panel: Instrumental throughput, calculated as the product of the detector QE and the total instrumental efficiency defined in Eq.~\ref{eq:efficiency}.}
    \label{fig:throughput}
\end{figure}
The stellar spectrum convolved at the resolution of Mauve is converted from flux to counts, in units of digital numbers (DN). First, the various payload efficiencies are considered, as illustrated in Fig.~\ref{fig:efficiencies}. These include:  
\begin{itemize}
    \item grating efficiency ($GR$),  
    \item fibre efficiency ($FB$),  
    \item coupling losses ($CL$),  
    \item two telescope mirrors ($MT$),  
    \item two spectrograph mirrors ($MS$),  
    \item spectrometer DCL efficiency ($DCL$),  
    \item spectrometer window efficiency ($WD$),  
    \item light loss ($LL$).  
\end{itemize} 
The overall instrumental efficiency is determined by multiplying these factors together:  
\begin{equation}
    \text{efficiency} = GR \cdot FB \cdot (1-CL)^4 \cdot MT^2 \cdot MS^2 \cdot DCL \cdot WD \cdot (1-LL) \, .
\label{eq:efficiency}
\end{equation}  
The coupling loss is currently defined as a 4\% loss at each fibre connection and at each boundary of the spectrometer window.

The instrumental efficiency is then combined with the detector quantum efficiency ($QE$), expressed in units of $\text{A/W}$, to determine the system's overall throughput. Both the quantum efficiency and the resulting throughput are illustrated in Fig.~\ref{fig:throughput}.

We combine the result thus obtained with the telescope's collecting area \( A \), the wavelength range per pixel \( \Delta x_{\lambda} \), and the exposure time \( t_{\text{exp}} \), which represents the duration of the observation. The input flux (in units of W/cm\(^2\) nm) is combined with the factors described above to convert the data in units of [A·s]. Subsequently, a series of conversion factors are applied to transform the data in units of counts. 
The conversion assumes that the 16-bit dynamic range is matched to the voltage dynamic range in the detector readout. According to the detector data sheet, the optimal voltage range is approximately \( 2.2 \, \text{V} \). The total counts are calculated as:  
\begin{equation}
    \text{counts} \ [\text{DN}] = F_i \cdot A \cdot \text{efficiency} \cdot QE \cdot \Delta x_{\lambda} \cdot t_\text{exp} \cdot \text{conversion factors} \, ,
\end{equation}  
where \( F_i \) represents the flux of each data point in the input spectrum.

\subsection{Adding astrophysical and instrumental noise sources}
\begin{figure}
    \centering
    \includegraphics[width=1\columnwidth]{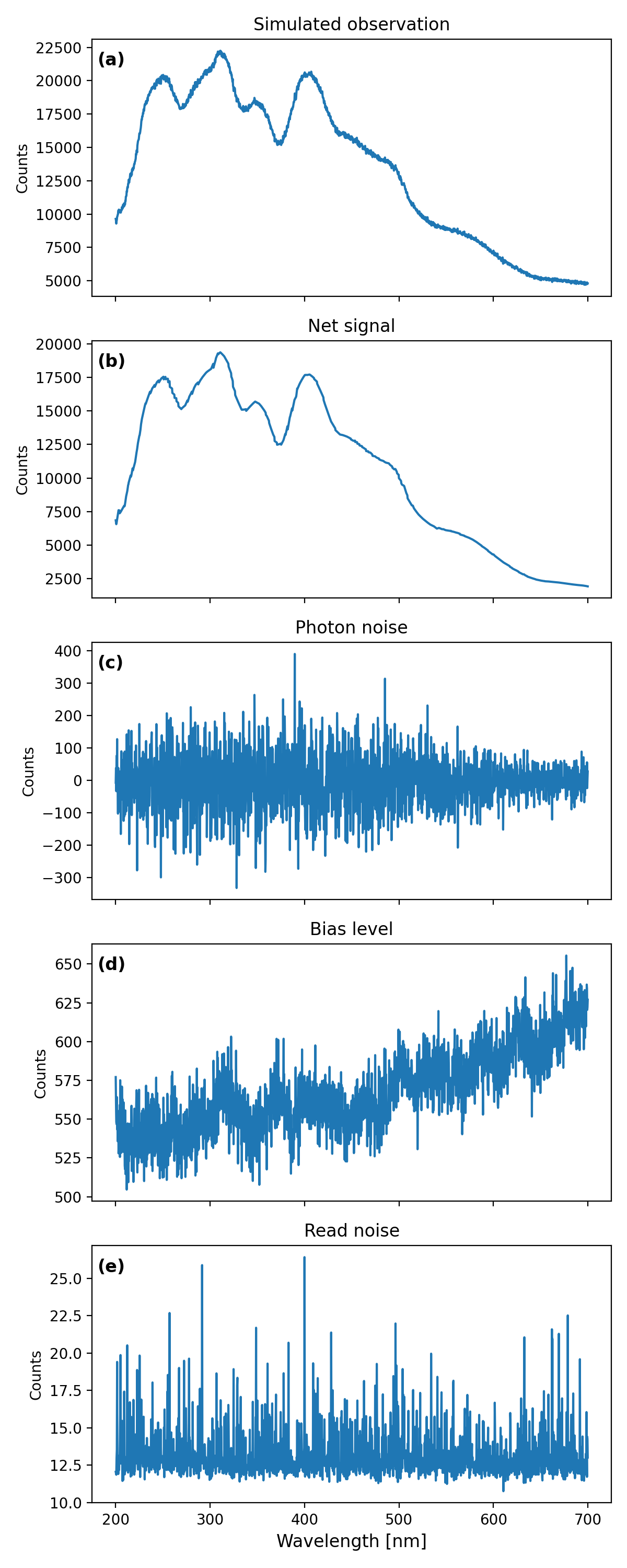}
    \caption{Simulation of HD 37202 using an exposure time of 5 s. The final simulated observation and its components are shown as well. Panel (a): output Mauve spectrum as simulated by the software. Panel (b): net simulated signal source. Panel (c): photon noise. Panel (d): bias level. Panel (e): read noise.}
    \label{fig:output}
\end{figure}
\begin{figure*}
    \centering
    \includegraphics[width=1\linewidth]{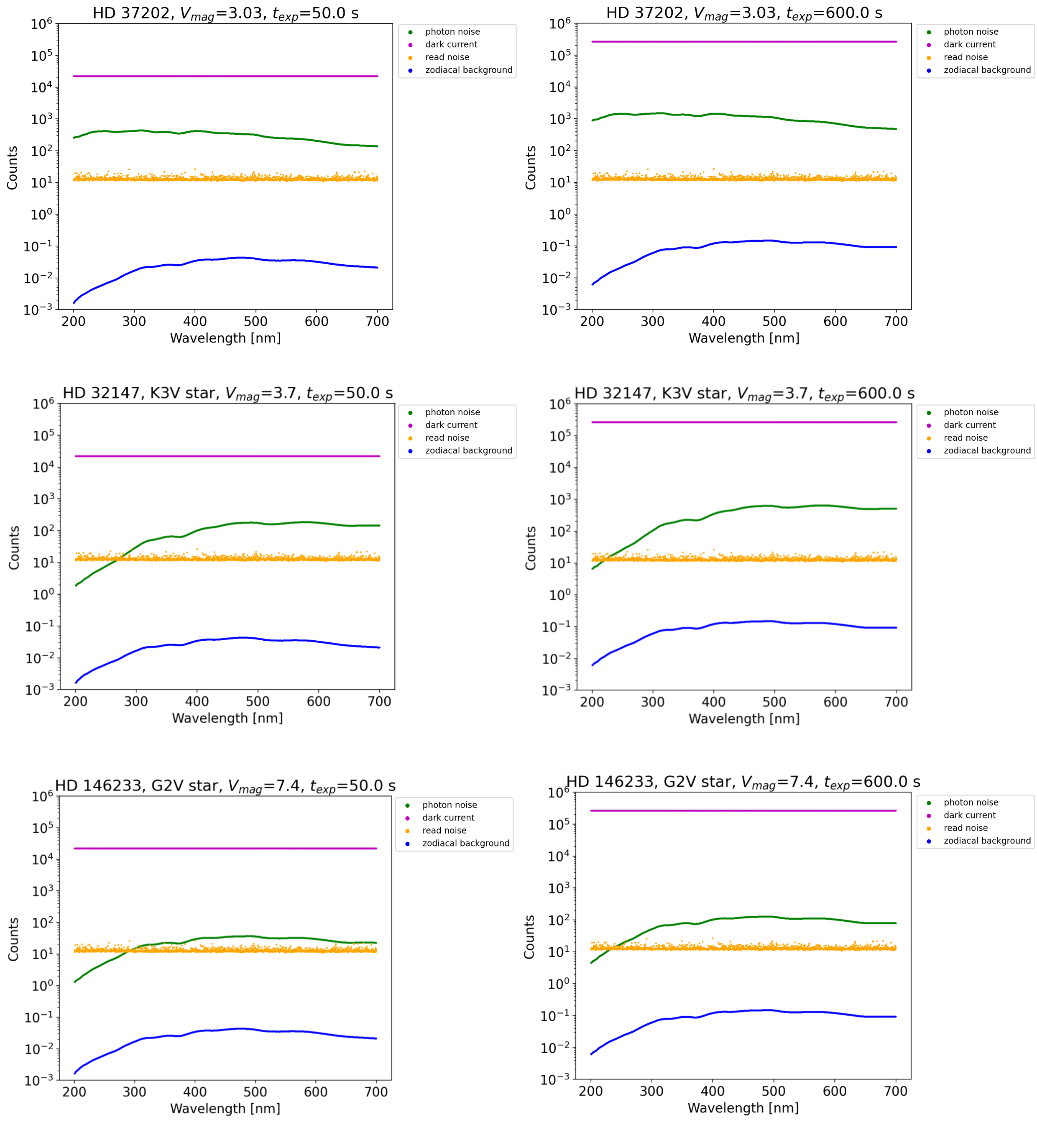}
    \caption{Diagrams of the noise budget illustrating the contributions from each payload noise source, presented for different target types and at two distinct exposure durations: 50 s and 600 s. We considered the Be star HD 37202 and the K3V star HD 32147 which are bright targets for the mission, and HD 146233, a fainter G2V star. The plots show the source photon noise in green, detector dark current in magenta, the read noise in yellow and the photon noise from the zodiacal background in blue.}
    \label{fig:noise_budget}
\end{figure*}
The result from the previous step represents the net source signal, excluding all noise sources except for the zodiacal background, which we include in the target source signal. For realistic simulations, additional noise components such as photon noise, bias, and dark current must be considered.  

Photon noise, which follows Poisson statistics, is modelled as the square root of the net signal in units of \( e^- \). To simulate the distribution of photons striking the detector, we include a random component to the photon noise which is drawn from a Gaussian distribution with $\mu$=0 and $\sigma$=$\sqrt{\text{net signal} [e^-]}$. Finally, to align the photon noise with the units of the source signal, photon noise is converted from \( e^- \) to counts.

Dark current was determined by empirically fitting a series of measurements taken at different temperatures, and it was found to scale according to the following relationship 
\begin{equation}
    I_{\text{d}}(T) = 105.94 \cdot e^{0.0684 \cdot T} \ ,
\end{equation}
where $T$ is the temperature of the detector. Based on this equation, the dark current is 445 counts/s at 21°C as reported in Table~\ref{tab:performance}. This noise component is added by drawing random samples from a Poisson distribution with expectation value equal to $I_{\text{d}}(T)$ multiplied by the exposure time. 

The bias is an offset applied to the ADC (Analogue-to-Digital Converter) to prevent negative counts during readout. The bias was measured in the lab and found to have an average value of 600 counts per pixel across the detector (Fig.~\ref{fig:output} (d)). 

Finally, readout noise is incorporated into the simulation (Fig.~\ref{fig:output} (e)). It was estimated as the standard deviation of the bias for each pixel, determined from a set of 556 bias frames.

The combination of the average dark current value and bias is what we define as the noise floor.

To illustrate how the various noise sources scale with target magnitude, spectral type and exposure time, Fig.~\ref{fig:noise_budget} shows the noise budget for the Be star HD 37202 and the flaring solar-type stars discussed in Sec.~\ref{sec:flaring_stars}.

The MauveSim user interface in Stardrive prompts the user to specify the exposure time for which the simulation should be generated. The interface also alerts the user if the selected settings (exposure time and source magnitude) result in detector saturation. The software outputs a one-dimensional raw data spectrum, as illustrated in Fig.~\ref{fig:output} (a). Mauve is planned to be operated with a 5-second cadence, meaning that each observation will consist of a series of 5-second frames. If the user selects an exposure time that is not a multiple of 5 seconds, the software will automatically round the exposure time up to the nearest multiple of 5. In this frame-stacking configuration, the signal from each frame is combined linearly, while the noise is added in quadrature.

\subsection{Signal-to-noise calculation}
\begin{figure}
    \centering
    \includegraphics[width=\columnwidth]{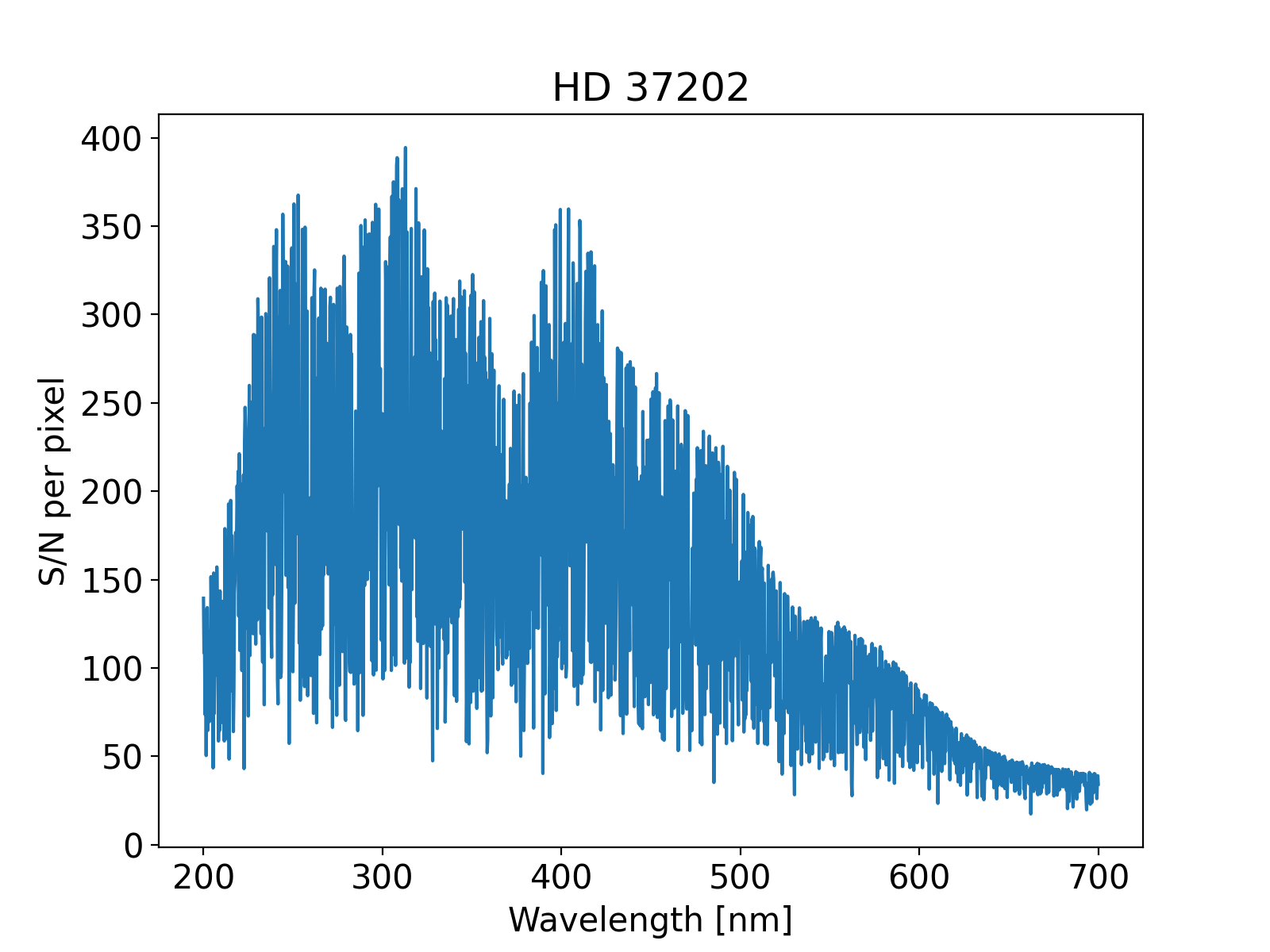}
    \caption{The S/N per pixel of the simulated data in Fig.~\ref{fig:output} (a).}
    \label{fig:snr}
\end{figure}
\begin{figure}
    \centering
    \includegraphics[width=1\linewidth]{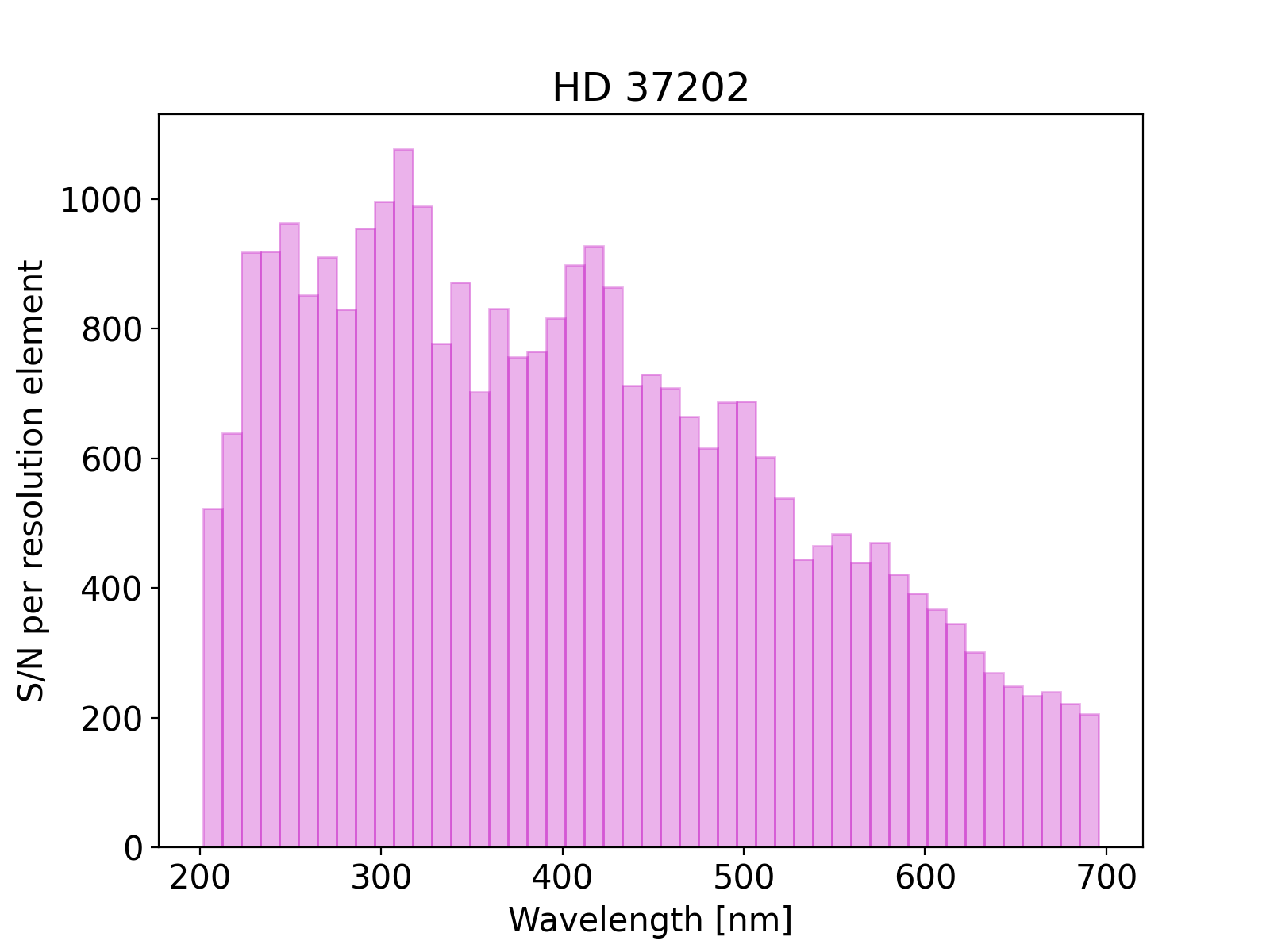}
    \caption{S/N per resolution element obtained on the simulation of HD 37202 using an exposure time of 5 s.}
    \label{fig:snr_binned}
\end{figure}
The S/N for each pixel, is calculated as follows
\begin{equation}
    S/N \ |_{\text{pix}} = \frac{S}{\sqrt{\sigma_S^2 + \sigma_R^2 + \sigma_D^2}} \ ,
\end{equation}
where $S$ is the net signal in units of counts (Fig.~\ref{fig:output} (b)); $\sigma_S$ is the photon noise ($\sqrt{S}$); $\sigma_R$ is the readout noise (equal to approximately 12 counts); $\sigma_D$ is the dark current noise (equal to the sqrt of the dark current, i.e. $\sqrt{\sim 2220 \ \text{counts}}$ for the example in Fig.~\ref{fig:output}). Fig.~\ref{fig:snr} displays the S/N per pixel of HD 37202 for a 5 s integration. 

Alternatively, the S/N can be calculated per resolution element (Fig.~\ref{fig:snr_binned}). This involves summing the signal from all pixels within the resolution element (10.5 nm, or approximately 40 pixels) linearly, while combining the corresponding noise components in quadrature:
\begin{equation}
    S/N \ |_{\text{bin}} = \frac{\sum_{i} S_i}{\sqrt{\sum_{i} N_i^2}}
\end{equation}
where $i = \{ k \mid \lambda_k \in [\lambda_{\text{lower}}, \lambda_{\text{upper}}) \}$, $\lambda_{\text{lower}}=\lambda_{\text{current}}, \lambda_{\text{upper}}=\lambda_{\text{lower}} + \Delta \lambda$, $\Delta \lambda = 10.5$ nm.

\section{Science case studies}
\label{sec:science_cases}
The “sweet spot” for Mauve’s science is spectrophometric monitoring of bright stellar sources in the near UV and visible wavelengths (200-700 nm) to search for variability on timescales ranging from minutes to weeks and months.

Mauve’s performance allows to effectively study flares in nearby active stars with observations not possible with any facility to date. For this reason, we will analyse this science case in detail, as it is Mauve’s “flagship science”. At the same time, other science areas can benefit from the same type of observations (spectrophotometric variability across the mid-near UV and visible bands). These include the
study of variability in young, active stars such as Herbig Ae/Be stars, the study of variability in (more evolved) Be stars, monitoring of active binaries and characterisation of blue straggler stars.

\subsection{Flaring solar-type stars}
\label{sec:flaring_stars}
Our own Sun is a very quiet star compared to similar-type stars in the Galaxy. The advent of white light monitoring space telescopes dedicated to exoplanetary transit searches (e.g., Corot, Kepler, TESS) has shown that even in visible light many stars emit significant flares, much larger than anything observed in the Sun \citep{2012Natur.485..478M}. However, during a flaring event, each wavelength probes a distinct region of the solar atmosphere \citep{2019ApJ...878..135K}. Consequently, while the white light curves provided by Kepler and TESS enable valuable statistical analyses of flare frequency across large stellar samples, they offer only limited potential for exploring the underlying physics of these events \citep{2014ApJ...797..121H, 2016ApJ...829...23D, 2020AJ....159...60G}.

A broader wavelength coverage of flaring events achievable with Mauve can enable the different spectral components to be disentangled. While with e.g. Kepler and TESS data, one sees the tail of the emission in white light, observations with Mauve would allow to test models of the emission from the heated stellar photosphere and lower chromosphere; different models of this exist, which scientists are interested to compare against actual observations \citep{2024LRSP...21....1K}. In particular, flares observed in white light are probably the result of light emitted from the heated, optically thick photosphere and lower chromosphere, that are likely to reach temperatures of 10,000 K or so, peaking in the near-UV band \citep{2013ApJS..207...15K}. By observing only the white light, one effectively sees the peak of the iceberg, while missing the bulk of it. With its wide wavelength coverage and the capacity to monitor specific sources for long periods, Mauve is ideal for studying individual large flares in detail, advancing the understanding of their physical mechanisms. Many exoplanets have been detected around active stars, and the UV emission from the host star dramatically influences the planetary evolution \citep{2010AsBio..10..751S, 2018haex.bookE..73S, 2020ApJ...902..115H}. Understanding the nature of this UV emission is crucial to gaining insights into broader topics, such as habitability.

We know from Kepler and TESS data that some stars similar to the Sun undergo “super flares”, events in which their white light increases by several percent (events up to some 25\% flux increase in the Kepler band have been observed, e.g. \cite{2013ApJS..209....5S} for the star KIC 12354328). If this flux rise is indeed produced by a heated patch of the photosphere, the increase in the blue and even more in the UV will be much larger, creating a “sweet spot” for Mauve observations. To demonstrate Mauve's capabilities under different stellar brightness scenarios, we selected two example stars and used MauveSim to generate simulated observations. The example stars are a bright K3V star with $V_\text{mag}$=3.7 and a G2V star with a visual magnitude of 7.4.

\subsubsection{K3V star}
\begin{figure}
    \centering
    \includegraphics[width=1\linewidth]{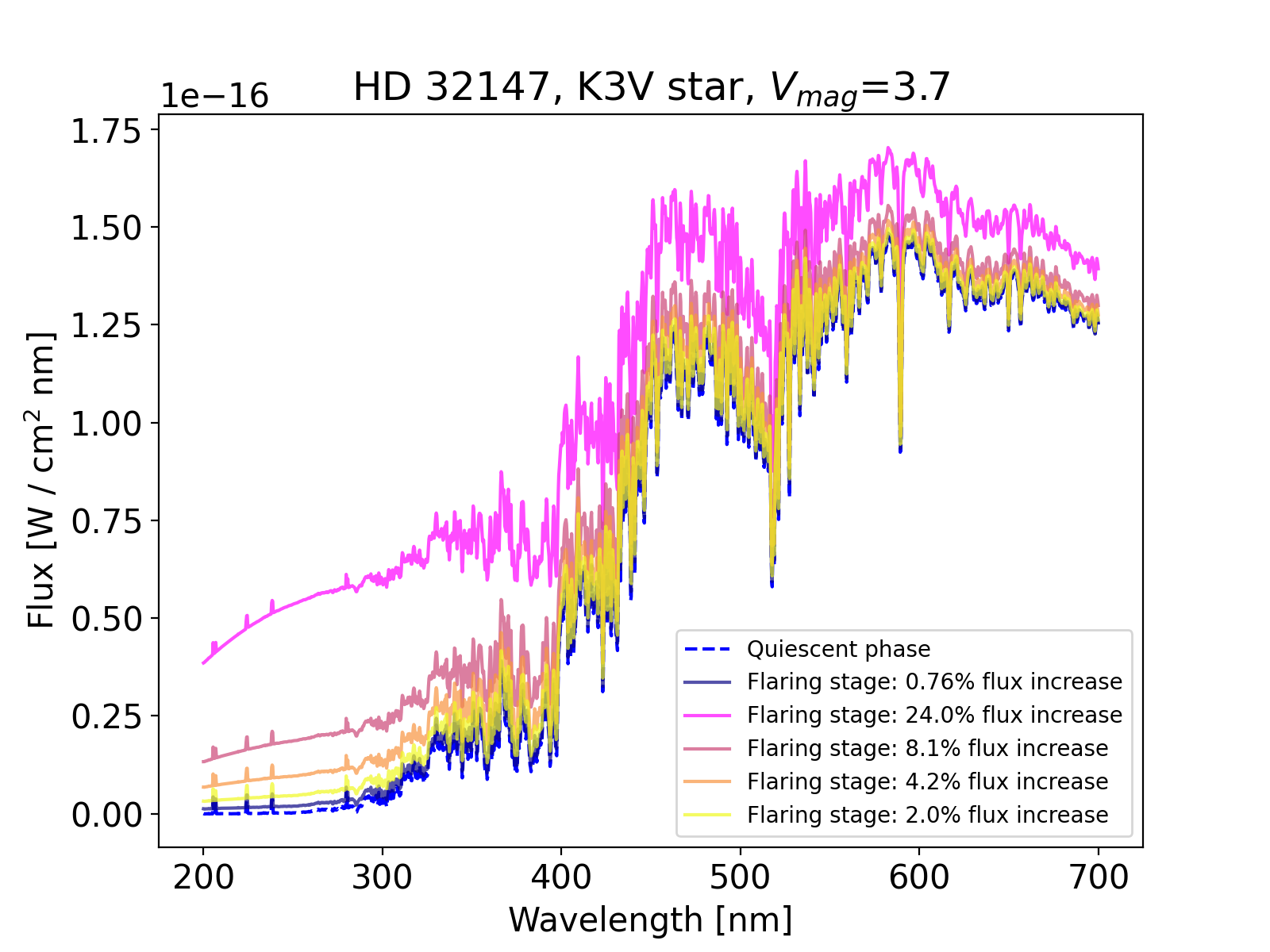}
    \caption{Example input spectra of a K3V star at different flaring stages. The quiescent spectrum is displayed in a dashed blue line. The flaring spectra contain a blackbody component with a temperature of 10,000 K that becomes increasingly more pronounced as the flare approaches its peak.}
    \label{fig:input_epsilon}
\end{figure}
\begin{figure}
    \centering
    \includegraphics[width=1\linewidth]{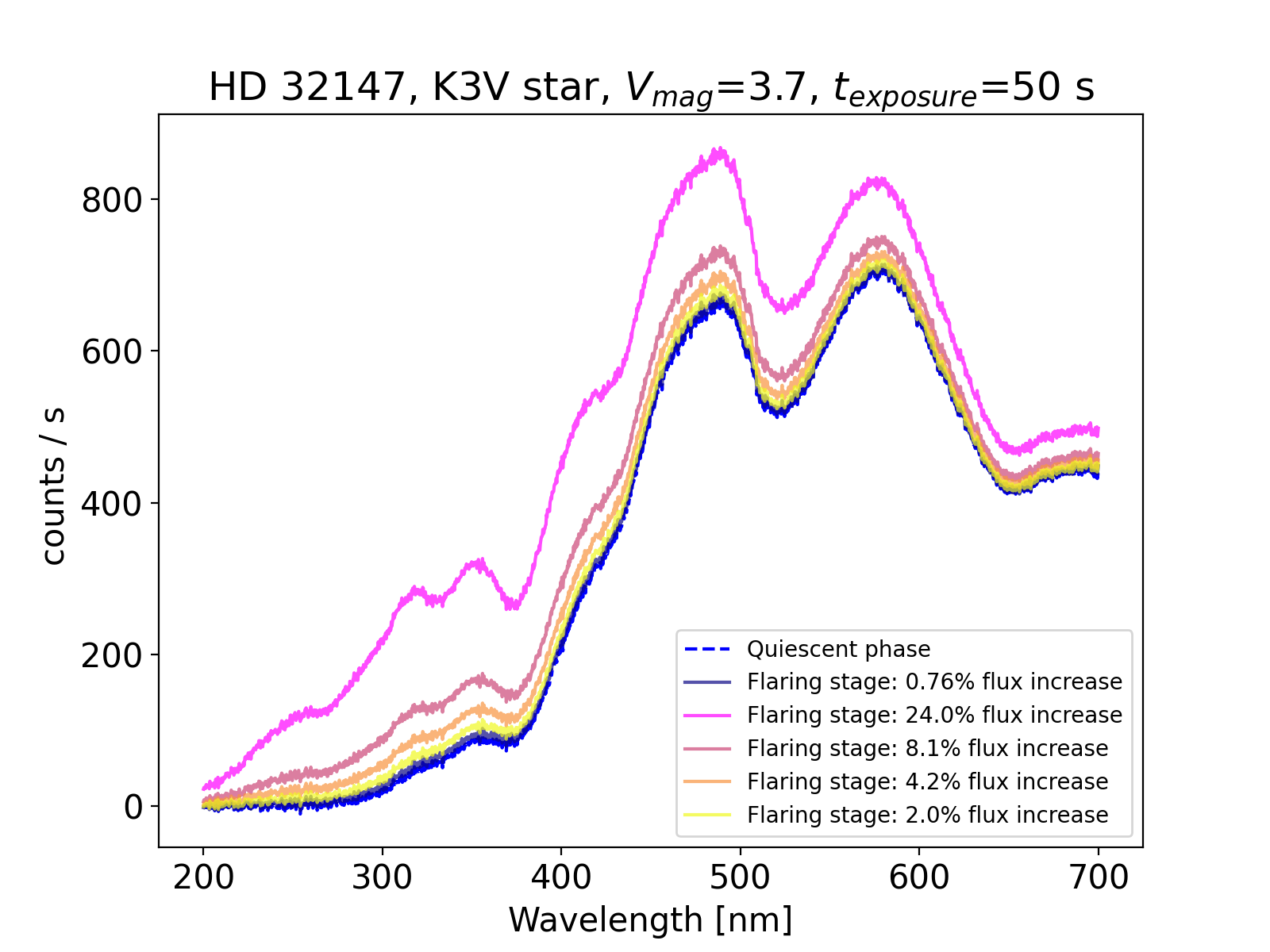}
    \caption{Simulated spectra in units of counts per second of a K3V star at different flaring stages, as observed by Mauve. These spectra, generated with MauveSim, have been background-subtracted, i.e. the average noise floor has been removed from each data point and the resulting values divided by the exposure time.}
    \label{fig:output_epsilon}
\end{figure}
The input spectra used to simulate the flare on the K3V star are shown in Fig.~\ref{fig:input_epsilon}. These display the stellar spectrum at different flaring stages, from the quiescent phase indicated with a dashed blue line, to the flare peak in magenta, and at varying decaying stages (the rest of the spectra in between). The flaring spectra include an additional component to them, which is a simplistic model of a blackbody at 10,000 K that becomes increasingly stronger as we approach the peak flaring stage. By selecting an exposure time of 50 s, we simulated how Mauve would see such spectra assuming the current performance. The corresponding MauveSim simulated spectra are shown in Fig.~\ref{fig:output_epsilon}. The low resolution of the instrument does not allow to retrieve individual lines. A comparison of the flaring spectra with the quiescent phase using chi-squared analysis reveals that the early flaring stage and the decaying phase (0.76\% and 2\% flux increase) are statistically indistinguishable from the quiescent spectrum ($\bar{\chi}^2$ $\ll$ 1). Flare peak and mid-flare spectra (i.e. 24\% and 8.1\% flux increase) show strong, statistically significant deviations ($\bar{\chi}^2$ $\gg$ 1), particularly the flare peak, which returns the highest $\chi^2$ value. While the greatest spectral contrast is observed in the ultraviolet region, the capability to distinguish the various flaring phases will be crucial for scientists aiming to test models of flare physics in solar-type stars.

\subsubsection{G2V star}
\begin{figure}
    \centering
    \includegraphics[width=1\linewidth]{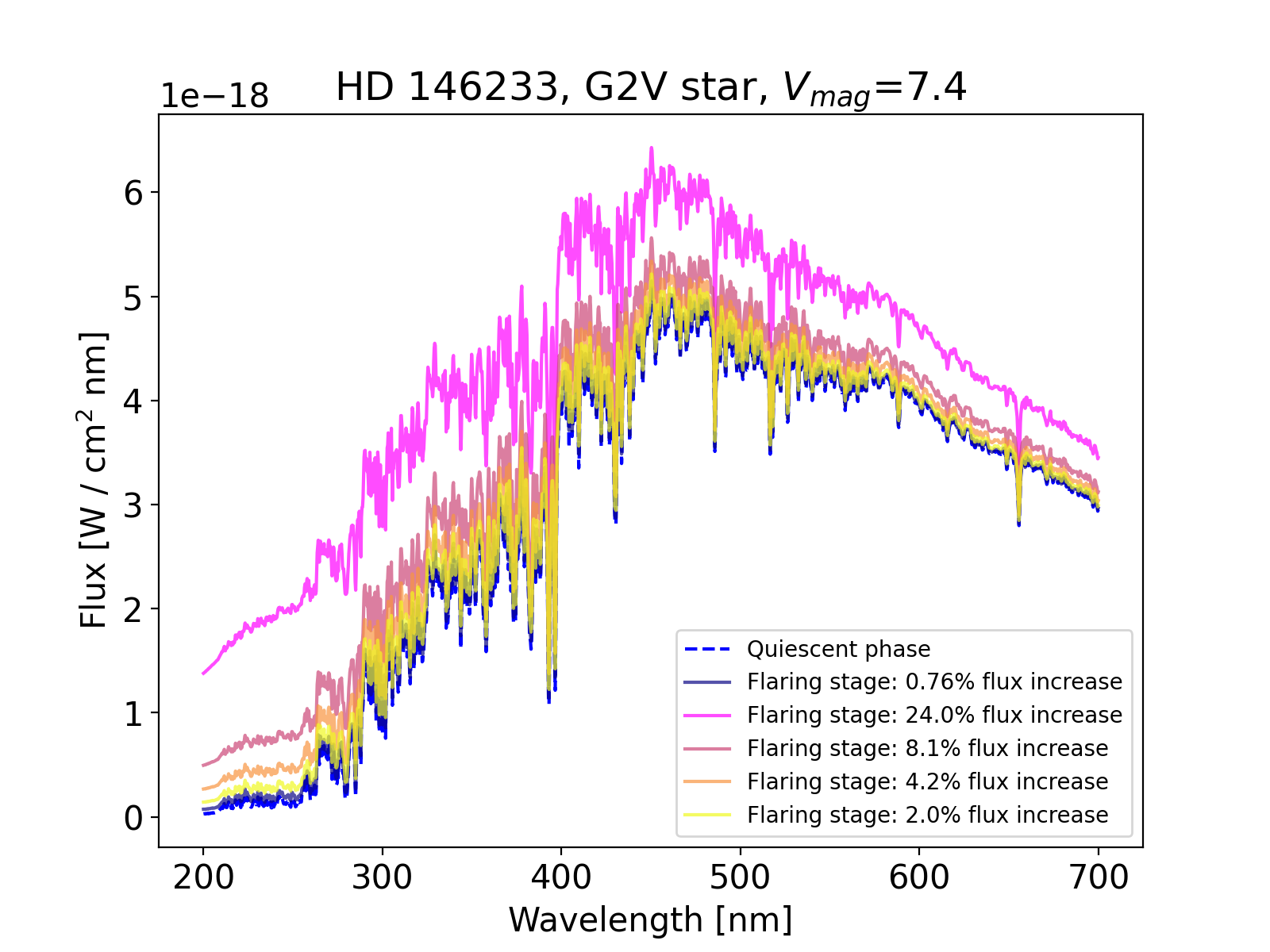}
    \caption{Example input spectra of a G2V star at different flaring stages. The quiescent spectrum is displayed in a dashed blue line. The flaring spectra contain a blackbody component with a temperature of 10,000 K that becomes increasingly more pronounced as the flare approaches its peak.}
    \label{fig:v889_input}
\end{figure}
\begin{figure}
    \centering
    \includegraphics[width=1\linewidth]{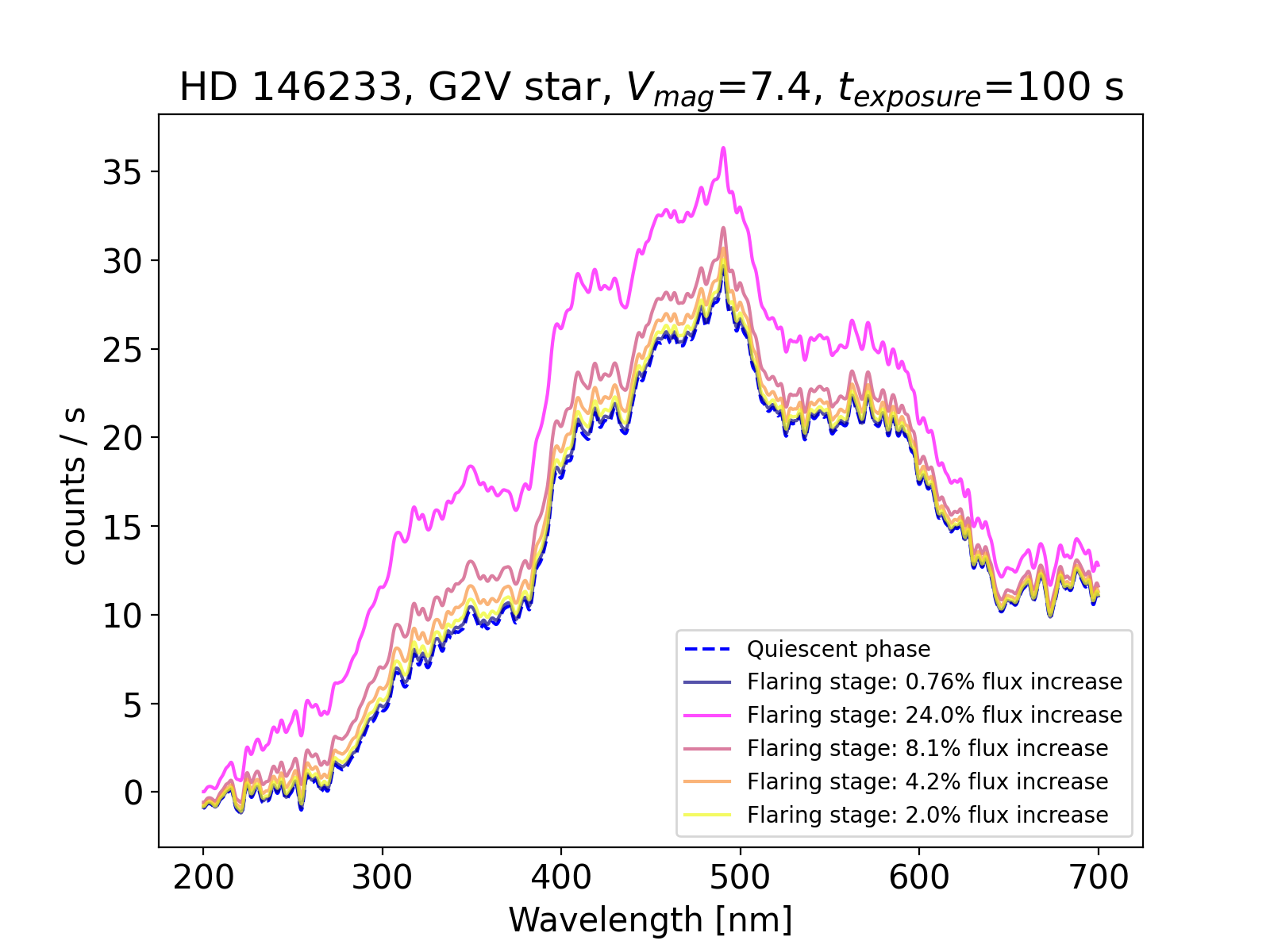}
    \caption{Simulated spectra of a G2V star at different flaring stages, as observed by Mauve. These spectra, generated with MauveSim, have been background-subtracted, i.e. the average noise floor has been removed from each data point and the resulting values divided by the exposure time. Additionally, a Gaussian kernel with a FWHM of 10.5 nm was applied to the spectra to reduce noise and help to visually distinguish the different flaring phases. This is an example near the limiting magnitude accessible to the Mauve’s flaring stars science case.}
    \label{fig:v889_output_base}
\end{figure}
For bright targets (V magnitude brighter than 6) Mauve can do excellent science on large flares, even in the presence of the high level of dark current. When going to fainter magnitudes, the dark current becomes the limiting factor, and, at magnitudes of 7.5 or so, even for a large flare such as one with a 25\% increase in the Kepler band, the noise level will limit the science that can be achieved.

To demonstrate this limitation, Figure~\ref{fig:v889_output_base} provides an example that is considered to be near the limiting magnitude for Mauve's flaring stars science case, i.e. the same large flare as for the K3V star, simulated for the fainter G2V stellar target in Fig.~\ref{fig:v889_input} ($V_\text{mag}$ = 7.4). A chi-squared comparison analysis between flaring and quiescent spectra reveals that the majority of the flaring spectra at the very beginning and end of the event are statistically indistinguishable from the quiescent phase. However, the flare peak stage creating a 24\% flux increase from quiescence returns a high chi-squared value ($\chi^2$ = 6159.88, $\bar{\chi}^2$ = 3.75), indicating a significant deviation from the quiescent state. Similarly, the flaring stage with a 8.1\% flux increase displays a moderate discrepancy from the quiescent phase ($\chi^2$ = 1078.36, $\bar{\chi}^2$ = 0.66). For a stellar target of this magnitude, it would be possible to distinguish the flare peak from quiescence. However, identifying a flaring event in the time series may be challenging unless the event is particularly energetic.

\section{Conclusions}
At Blue Skies Space we developed MauveSim, an in-house instrument simulator to support the Mauve mission. This software generates simulated observations by requiring an input spectrum representative of the target source, which can either be an observed spectrum from other facilities or a synthetic model spectrum. The input spectrum is processed through the instrument response, incorporating its spectral resolution, optical element efficiencies, and detector response. Astrophysical and instrumental noise sources are then added to the net source signal. Users can specify the exposure time, and the software produces a simulated raw spectrum as output. This tool is crucial in helping members of the Mauve science team refine their science cases and evaluate the feasibility of proposed observations. The software is accessible to all scientists who currently are or will be involved in the mission. Scientists not yet part of the Mauve survey are encouraged to explore the software by contacting BSSL by email at \url{info@bssl.space}.

\section*{Acknowledgements}
The Mauve project has received funding from the European Union’s Horizon Europe research and innovation programme under grant agreement No. 101082738. We thank ISISPACE, MediaLario and C3S for their partnership with BSSL in constructing and delivering Mauve. We also thank the anonymous referees for providing helpful comments that enhanced the quality of the manuscript.

\section*{Data Availability}
Data utilised as input to perform the simulations are publicly available in the HST/STIS Stellar Spectral Library at \url{http://astro.wsu.edu/hststarlib/}. 




\bibliographystyle{rasti}
\bibliography{bibliography} 





\bsp	
\label{lastpage}
\end{document}